\pdfoutput=1

\documentclass[twocolumn]{RAA}
\usepackage{graphicx,times}             
\usepackage{natbib}
\usepackage{amssymb,amsmath}
\usepackage{changes}

\bibpunct{
(}{)}{;}{a}{}{,}

\usepackage[pagebackref=true]{hyperref}

\def\kms{km s$^{-1}$}
\def\cms{cm$^{-2}$}

\def\s~{$\sim$}

\def\HI{H\,{\small{I}} }

\newcommand{\HIFAST}{\texttt{HiFAST} \ }

\begin{document}
\title{ Observation of HI around three satellite galaxies of the M31 with the FAST: Andromeda II, NGC 205, and NGC 185 }
  

   \volnopage{Vol.0 (20xx) No.0, 000--000}      
   \setcounter{page}{1}          

   \author{Ziming Liu \inst{1,2}
   \and Jie Wang \inst{1,2}
   \and Yingjie Jing \inst{1} 
   \and Chen Xu\inst{1,2}
   \and Tiantian Liang\inst{1,2}
   \and Qingze Chen \inst{1,2}
   \and Zerui Liu \inst{1,2}
   \and Zhipeng Hou \inst{1,2}
   \and Yougang Wang \inst{1}
   }
   

   \institute{
    National Astronomical Observatories, Chinese Academy of Sciences, Beijing 100101, China; {\it jie.wang@nao.cas.cn, zmliu@nao.cas.cn}\\
        \and
    School of Astronomy and Space Science, University of Chinese Academy of Sciences, Beijing 100049, China;
    \\
\vs\no
   {\small Received 2024 April 15; accepted 2024 May 7}}

\abstract{
With the exceptional sensitivity of the Five-hundred-meter Aperture Spherical radio Telescope (FAST), we conducted observations of the neutral hydrogen (HI) in the circumgalactic medium of Andromeda's (M31) satellite galaxies, specifically Andromeda II, NGC 205, and NGC 185. Initially, three drift scans were executed for these satellites, with a detection limit of $4\times10^{18}$ \cms ( approximately $1.88\times10^3 M_{\odot}$ of \HI mass), followed by a more in-depth scan of a specific region. We discovered a C-shaped HI arc structure sharing a position and line-of-sight velocity similar to a stellar ring structure around Andromeda II, hinting at a potential connection with Andromeda II. In the context of NGC 205, we identified two mass concentrations in the northeast direction, which could be indicative of tidal streams resulting from the interaction between this galaxy and M31. These new lumps discovered could be very helpful in solving the missing interstellar medium (ISM) problem for NGC 205. Observations regarding NGC 185 are consistent with previous studies, and we did not detect any additional HI material around this galaxy. These observational results enhance our understanding of the evolution of these satellite galaxies and provide insight into their historical interactions with the M31 galaxy.
\keywords{ISM: structure---local interstellar matter---galaxies: interactions---galaxies: ISM---Local Group---radio lines: ISM---radio lines: galaxies}
}
   \authorrunning{Z.M. Liu et al.}
   \titlerunning{HI Around Three Satellite Galaxies of M31}
   \maketitle

\section{Introduction}
\label{sec:intro}
The gaseous component of the universe plays a crucial role in the composition of galaxies. Understanding its distribution and dynamics is essential for uncovering the formation, evolution, and interaction of galaxies. Atomic neutral hydrogen (\HI) is particularly significant, serving not only as a principal element of the ISM and a fundamental ingredient for star formation, but also as a useful tracer in the CGM to track the interaction history of two galaxies. Substantial progress has been made in both observational and theoretical research on the ISM within the Milky Way and its Local Group companions, shedding light on the unique characteristics of dwarf galaxies, which may exhibit ISM properties distinct from those of the larger spiral galaxies with their spiral density waves and bars \citep{1974Natur.252..111E,2000ApJ...541..675B,2003AJ....125.1926G,2009ApJ...696..385G,2014ApJ...795L...5S}. Furthermore, research on the HI component in the CGM and IGM has uncovered that the galaxies are in a phase of interaction, even though they appear separated and unrelated based on observations in the optical bands. This offers crucial proof of their interaction and an essential understanding of their evolutionary paths.

The Andromeda Galaxy (M31) is the closest typical galaxy to us, and studying the HI content in its satellite galaxies is expected to offer valuable insights into intergalactic interactions and hydrodynamical behaviors \citep{1998ARA&A..36..435M,2005ASPC..331..113T,2012AJ....144....4M,2019ARA&A..57..375S}. Understanding the distribution and dynamics of gaseous material is crucial for unraveling the evolutionary past of interactions between M31 and its satellites. Nevertheless, within the virial radius of M31, most dwarf galaxies exhibit a dearth of gas, whereas, beyond this threshold, gas predominates \citep{2012AJ....144....4M,2021ApJ...913...53P}. Henceforth, NGC 185 and NGC 205, as two dwarf galaxies exhibiting substantial gas masses within the virial radius of M31, have attracted heightened scrutiny, notwithstanding the indeterminate origins of their gaseous content. \citep{2003AJ....125.1926G,2009ApJ...696..385G,2017MNRAS.465.3741D}. Additionally, indications from both \HI and other bands observations suggest the potential presence of interactions between NGC 205 and M31 \citep{1988A&A...200...21C,1991A&A...246..349B,2005AJ....130.2087D,2012MNRAS.423.2359D}. Furthermore, although no \HI radiation has been observed internally in Andromeda II, the second-largest satellite galaxy of M31, its position overlaps with the \HI gas clouds within the Milky Way, suggesting the possibility of obscuration. Optical observations of Andromeda II reveal remnants of past mergers, with peripheral stellar streams possibly originating from previous merger events\citep{2012ApJ...758..124H,amorisco2014remnant,2017MNRAS.469.4999D}. The merger history of this satellite galaxy has hitherto been elucidated only through numerical simulations of stellar observational results \citep{2014MNRAS.445L...6L,2017MNRAS.464.2717F,2019IAUS..344...62E}. 

With the high sensitivity of the FAST, it will be very interesting to revisit these three satellite galaxies, especially the observation of the HI in the CGM medium of these galaxies will provide important information on their interaction with the host M31 galaxy. In this paper, we present a very deep \HI observation of the CGM around these three satellites. In particular, we checked the HI content in the CGM of these three satellites to investigate the diffuse \HI component around these galaxies. Section 2 describes the sources and methods of data acquisition and processing. Section 3 delves into the \HI distribution and dynamics of these three distinct satellite galaxies. Finally, Section 4 succinctly summarizes the main findings of this study. Throughout this paper, we adopt distances of 0.8, 0.66, and 0.65 Mpc to NGC 205, NGC 185, and Andromeda II, \citep{2018MNRAS.479.4136K} respectively.

\section{DATA} \label{sec:data}

\subsection{Observation} \label{subsec:observation}

Utilizing the 19-beam receiver of FAST, we are able to conduct simultaneous observations of multiple areas, significantly enhancing the efficiency of observing larger target areas. To effectively detect the diffuse \HI components surrounding the three dwarf galaxies, we set the size of the target fields to be at least five times the angular diameter of the dwarf galaxies in optical images. Employing a combination of the DriftWithAngle and MultibeamOTF modes \citep{2019SCPMA..6259502J,2020RAA....20...64J}, we observed our target fields, aiming to attain the desired signal-to-noise ratio. Both methods capture the targets through a scanning way. In the DriftWithAngle mode, the FAST receiver rotates by $23.4^\circ$ to align with the target's declination, maintaining a static attitude. This mode relies on the Earth's rotation to scan the sky at a speed of $15^{\prime\prime}/s$. It ensures a constant zenith angle for the telescope during observation, thereby aiding in maintaining stable gain. The MultibeamOTF mode scans designated rectangular areas along either the right ascension (with a receiver rotation angle of $53.4^\circ$) or declination (with a receiver rotation angle of $23.4^\circ$) directions, with a scanning interval of $21^\prime.66$. We set the scanning speed for MultibeamOTF observation to $1^{\prime\prime}/s$ to achieve optimal integration time. Each MultibeamOTF observation lasts approximately 150 minutes. During observations, the FAST spectral-line backend was configured in the W+N mode, with a frequency channel width of 7.62939 kHz and a corresponding velocity resolution of 1.6 \kms at z = 0. A high-mode noise diode (with the amplitude of $\sim 11$ K) is periodically injected to calibrate the \HI spectral data. We use the gain parameters of degrees per flux unit (DPFU) for 19 beams in  Liu et al. (2024, in prep.) to calibrate the flux density of the observation data. 

For each of the three dwarf galaxies, we conducted three separate observations using the DriftWithAngle mode to acquire the foundational dataset. The $3\sigma$ column density of our foundational dataset reaches $6.5 \times 10^{18}$ \cms, corresponding to a mass limit per beam of \HI of $2.78\times10^3 M_{\odot}$ at a distance of 0.65 Mpc. The data for NGC 185 is obtained directly from the primary dataset, with a total integration time of approximately 3 minutes. Furthermore, our dataset for NGC 205 is enriched by incorporating an additional MultibeamOTF mode deep scan, resulting in an enhanced column density sensitivity of $4\times10^{18}$ \cms ($1.88\times10^3 M_{\odot}$ of \HI mass). After stacking one additional observation, the integration time for NGC 205 reaches $\sim 2.3$ hours. In the Andromeda II field, the survey was intensified with two supplementary MultibeamOTF mode deep scans, with a total integration time of $\sim 5.3$ hours, yielding an improved \HI column density sensitivity of $6.5\times10^{17}$ \cms ($16.10 M_{\odot}$ of \HI mass).

\subsection{Data Reduction} \label{subsec:data}

The spectral data obtained from the observations were processed using the \HIFAST pipeline version 1.4, a software tool developed by \citet{2024arXiv240117364J} specifically for the analysis of \HI spectral observations conducted by FAST. \HIFAST provides a comprehensive array of functionalities crucial for the reduction of \HI spectral data, encompassing noise diode calibration, baseline subtraction, flux density calibration (Liu et al., 2024 in prep.), identification and removal of standing waves and radio frequency interference (RFI) (Xu et al., 2024 in prep.), and subsequent imaging \citep{2024arXiv240117364J}. These findings were meticulously cross-referenced with optical observational data to delineate satellite galaxies potentially containing \HI content. 

The processing of observation data for the three dwarf galaxy systems involved distinct methodologies. Given the relatively isolated nature of NGC 185, we directly employed the built-in methods of \HIFAST to individually image it, extracting the global \HI profile and calculating relevant parameters accordingly. In the case of NGC 205, its proximity to M31 and partial overlap with M31's line of sight on its western side necessitated a Gaussian fitting approach within the data cube to separate spectral lines in the overlapping regions, ensuring that only NGC 205's spectral components were retained in the final flux results. Andromeda II presents a particularly unique case among dwarf galaxy systems. Previous observations suggested the absence of \HI components within its interior. However, as the second-largest satellite galaxy of M31, it boasts an extensive optical radius ($9^\prime$), surrounded by numerous \HI gas structures. Most of these structures exhibit line-of-sight velocities inconsistent with those of Andromeda II, suggesting a more likely association with gas components within the Milky Way's halo. Nevertheless, the gas structure nearest to the northwest of Andromeda II displays velocities closely aligned with its optical components. Spectral data was extracted from this region using a plausible circular mask as a reference for analysis in our study. 

\begin{table*}[!h]
\centering
\caption[]{Position and \HI Parameters of three satellite galaxies}
\begin{tabular}{lccccc}
\hline\noalign{\smallskip}
Dwarf Galaxy & RA and DEC & $V_\odot$ (\rm{\kms}) & $D_\odot$ (\rm{Mpc}) & 1$\sigma$ (\rm{mJy}) &  1$\sigma$ (M$_\odot$) \\
\hline\noalign{\smallskip}
Andromeda II & 01\textsuperscript{h}16\textsuperscript{m}30\textsuperscript{s} +33\textdegree25$^\prime$09$^{\prime\prime}$ & -169.561 & 0.65 & 0.16 & 16.01\\ 
NGC 205 & 00\textsuperscript{h}40\textsuperscript{m}22\textsuperscript{s} +41\textdegree41$^\prime$07$^{\prime\prime}$ & -221.224 & 0.8 & 12.42 & $1.88\times10^3$\\
NGC 185 & 00\textsuperscript{h}38\textsuperscript{m}58\textsuperscript{s} +48\textdegree20$^\prime$15$^{\prime\prime}$ & -200.875 & 0.66 & 27.04 & $2.78\times10^3$\\
\hline\noalign{\smallskip}
\end{tabular}
\label{tab:para}
\end{table*}

\begin{table}[!h]
\centering
\caption[]{Comparison of \HI mass in $10^5$ M$_\odot$}
\begin{tabular}{lcccc}
\hline\noalign{\smallskip}
Dwarf Galaxy & dGM97 & dGM16 & dGM21 & dGM24 \\
\hline\noalign{\smallskip}
Andromeda II & * & $\textless$11 & $\textless$0.44$\pm$0.02 & 1.7$\pm$0.5 \\ 
NGC 205 & 4.3 & 3.8 & 4.0$\pm$0.3 & 4.2$\pm$0.3 \\
NGC 185 & 1.0 & 1.3 & 1.1$\pm$0.1 & 1.0$\pm$0.1 \\
\hline\noalign{\smallskip}
\end{tabular}
\label{tab:himass}
{\small
Where dGM97\citep{1997ApJ...476..127Y}, dGM16\citep{2016ApJ...824..151G}, dGM21\citep{2021ApJ...913...53P} and dGM24 (our finding) is the mass or detection limit for the three galaxies. }
\end{table}

\begin{figure*}[ht!]
\includegraphics[width=\textwidth, angle=0]{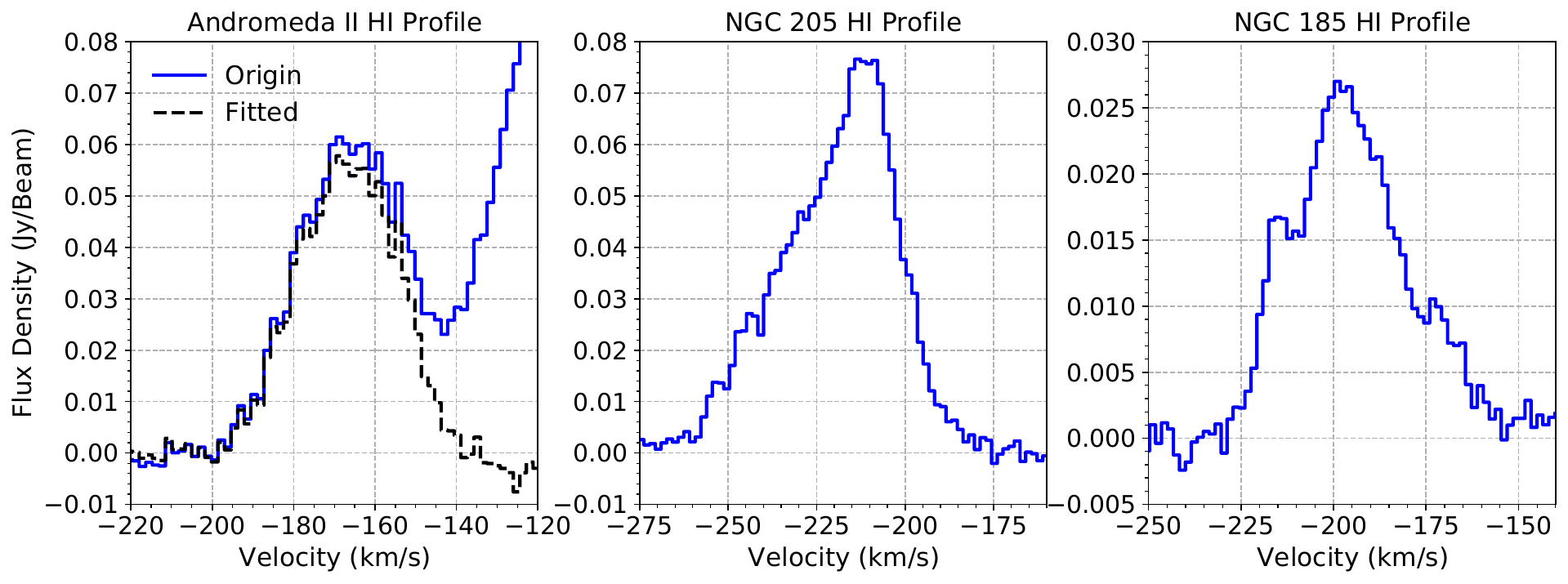}
\caption{The Global \HI profiles of Andromeda II, NGC 205, and NGC 185 (from left to right panel). Given that the ambient gas of And II is intermingled with that of the Milky Way, a single Gaussian component with a velocity of approximately -170 \kms is utilized to obtain the parameter of And II. }
\label{fig:HIProf}
\end{figure*}

\section{RESULTS} \label{sec:results}

In this section, we will provide an overview of these three galaxies first, followed by a detailed presentation of the results for each of them separately.

Figure~\ref{fig:HIProf} illustrates the overall hydrogen profiles of these three galaxies. The profiles exhibit a two-horn or three-peak structure, with one peak being significantly stronger than the others. This suggests the presence of multiple components in these galaxies, making their dynamic state somewhat intricate. For instance, in the case of NGC 185, there are three peaks: a prominent peak at $-200$ \kms, along with two weaker peaks at $-215$ \kms and $-175$ \kms. Similarly, NGC 205 shows a strong peak at $-215$ \kms and a very faint peak at $-245$ \kms. The profile of Andromeda II is challenging to discern due to contamination from Milky Way components. Only components within the velocity range of $-200$ \kms to $-140$ \kms are considered for fitting after excluding the surrounding masked regions, the fitted profile reveals a strong peak at $-169$ \kms. Further details on this source will be discussed in the subsequent subsection. 

Table~\ref{tab:para} provides an overview of the center velocity ($V_\odot$), distance ($D_\odot$)\citep{2018MNRAS.479.4136K} and characteristics related to the \HI component for the trio of satellite galaxies. The mass or detection thresholds for the mass of these three dwarf galaxies were previously outlined by \citet{1997ApJ...476..127Y} (hereafter dGM97), \citet{2009ApJ...696..385G} (erratum by \citet{2016ApJ...824..151G}, hereafter dGM16) and \citet{2021ApJ...913...53P} (hereafter dGM21). The \HI mass presented in our findings, referred to as dGM24 hereafter, is compared with theirs in Table \ref{tab:himass}. It is clear that our estimation of the \HI mass of Andromeda II falls within the range reported by \citet{2016ApJ...824..151G}, but exceeds that of \citet{2021ApJ...913...53P}.

The deviation in this value stems from our measurement of the \HI mass of Andromeda II, which encompasses the collective mass of its surrounding gas structures rather than exclusively the \HI mass confined within the dwarf galaxy itself. Consequently, it is plausible that it may exceed the upper limit of the \HI mass within the central region of the dwarf galaxy. It is worth mentioning that the RMS of \HI flux density in our dataset for Andromeda II is 0.16 mJy, notably lower than the value reported by \citet{2021ApJ...913...53P} (12 mJy). Meanwhile, we have not identified any isolated \HI structures within the central region of Andromeda II at -192 \kms. Furthermore, our determination of the \HI mass of NGC 185 concurs closely with several prior observations. In the case of NGC 205, our findings align more closely with the conclusions drawn by \citet{1997ApJ...476..127Y} and \citet{2021ApJ...913...53P}, and just 10 percent higher than the value reported by \citet{2016ApJ...824..151G}.

\subsection{Andromeda II} \label{subsec:andii-results}

\begin{figure*}
\centering 
\includegraphics[width=\textwidth, angle=0]{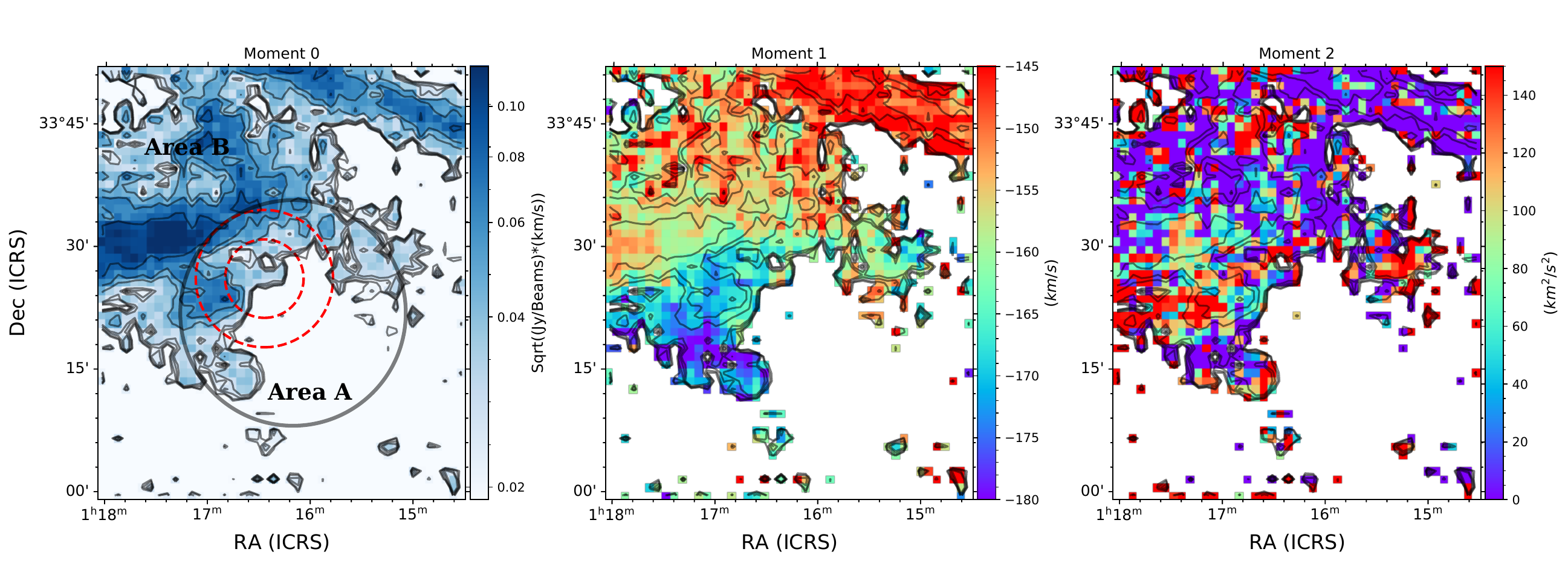}
\caption{The moment 0 (left), moment 1 (middle), moment 2 (right) depicting the region surrounding Andromeda II. The contours display the levels of \HI column density at 6.5, 7, 8, 9, 11, 15, 20, and 30 $\times 10^{17} $ \cms. A C-shape arc structure (black circle) is approximately 13.8 arc minutes far away from Andromeda II, opening towards the southeast. The bigger red dashed ring shows the optical range of Andromeda II and the region between the two red dashed lines is the stellar stream, discussed by \citet{amorisco2014remnant}.}
\label{fig: AndII-M012}
\end{figure*}

\begin{figure*}[ht!]
\includegraphics[width=\textwidth, angle=0]{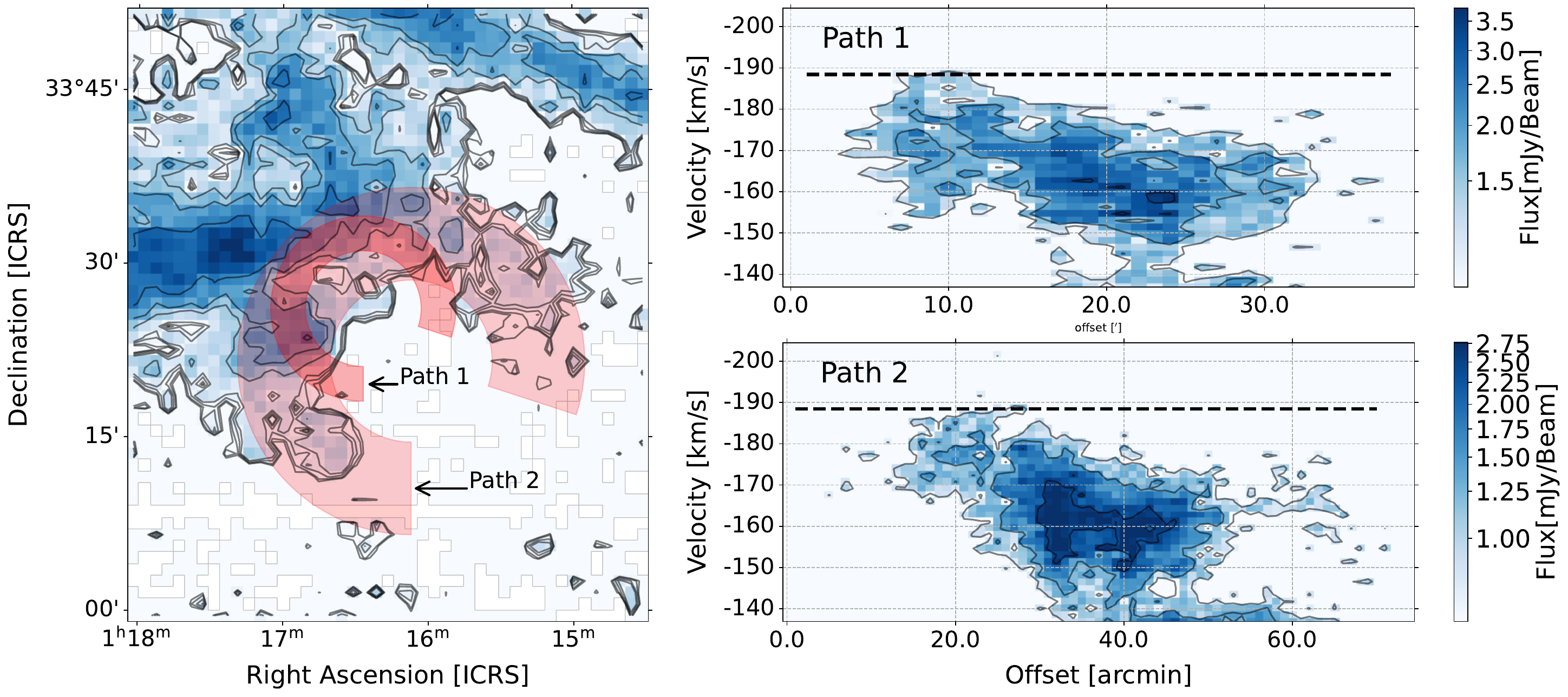}
\caption{Position and velocity diagram of the Andromeda II star stream region and gas arc region. The left panel displays the moment 0 of the region around Andromeda II, with contours identical to the previous Figure \ref{fig: AndII-M012}. The red arc area signifies the region utilized in the PV diagrams, and the black arrow indicates the starting position and direction of this area. The right panel presents the PV diagram corresponding to the left image respectively, where the black dashed line illustrates the averaged velocity distribution of the star stream. Contours highlight the 3, 4, and 5 $\sigma$ limits.}
\label{fig: AndII-ring-pv}
\end{figure*}

\begin{figure*}[ht!]
\includegraphics[width=\textwidth, angle=0]{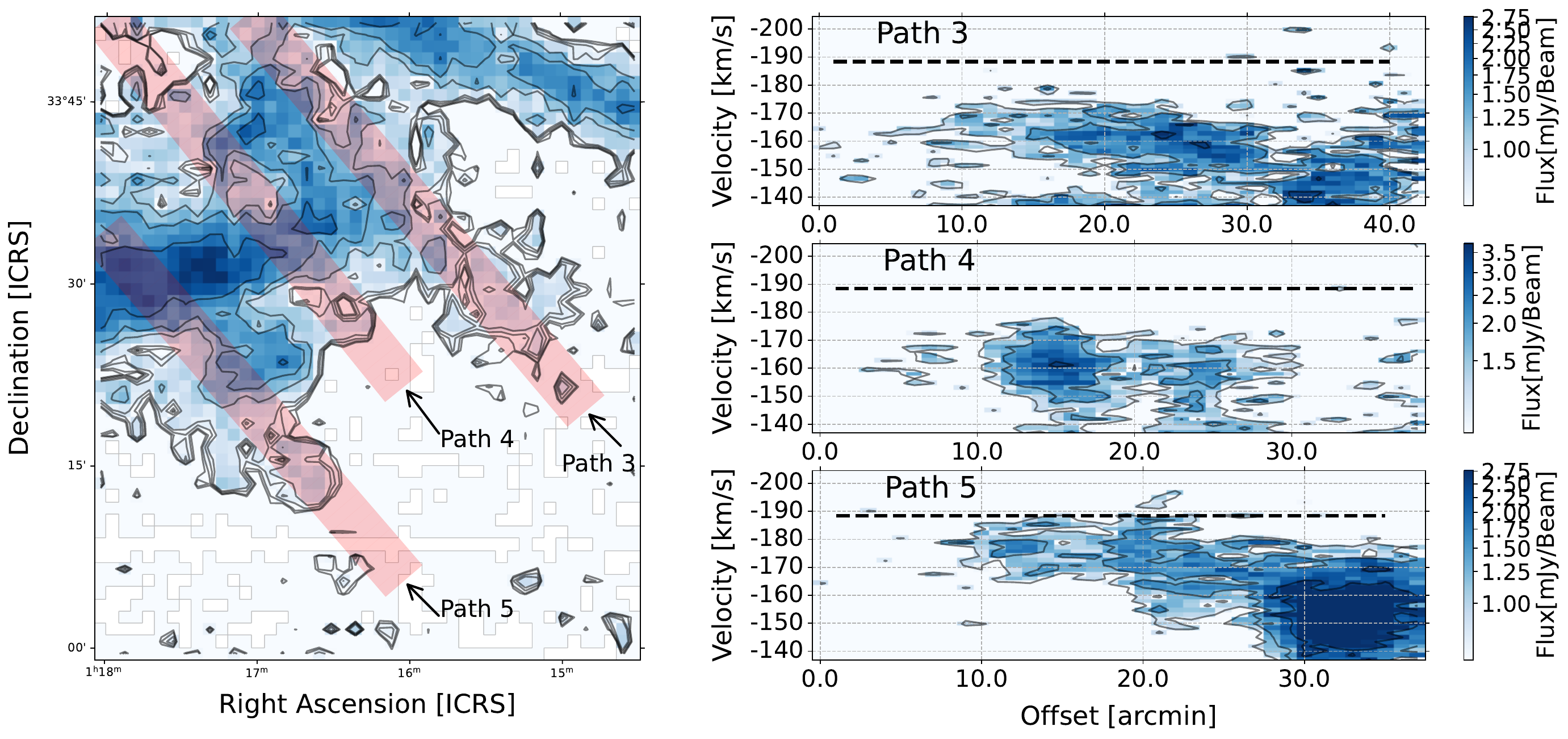}
\caption{Position and velocity diagram of the Andromeda II at other three directions. The left panel displays the moment 0 of the region around Andromeda II, with contours identical to the previous Figure \ref{fig: AndII-M012}. The red line areas signify the three regions utilized in the PV diagrams, and the black arrow indicates the starting position and direction of the region. The right panel presents the PV diagram corresponding to the left image respectively, where the black dashed line illustrates the averaged velocity distribution of the star stream. Contours also highlight the 3, 4, and 5 $\sigma$ limits.}
\label{fig: AndII-lines-pv}
\end{figure*}

Figure \ref{fig: AndII-M012} illustrates the map of the integrated \HI column density (Moment 0), as well as the Moment 1 and Moment 2 maps covering a $50^\prime \times 50^\prime$ area around Andromeda II. In the central part of Andromeda II, there is no clear distribution of \HI emission. However, about  13.8 arc min far away, there is a c-shape arc structure (named $Area A$, marked by a black circle). it appears in the north and west, with its opening facing southeast. Interestingly, this pattern closely resembles the outer stellar stream of Andromeda II previously identified by \citet{amorisco2014remnant}, showing similar orientations. The area containing the stellar stream is marked by red dashed lines in the left panel of Figure \ref{fig: AndII-M012}. 

Moreover, $Area A$ seems to be linked to a larger gaseous region labeled as $Area B$ towards the northwest. Examination of the Moment 1 map indicates a similar velocity trend for the gas in both $Area A$ and $Area B$, with velocities generally increasing from southwest to northeast. Additionally, analysis of the Moment 2 map reveals higher velocity dispersion along the western and northern boundaries of $Area A$, suggesting that the gas in these areas are separated objects. 

To investigate the potential connection between the \HI gas in $Area A$ and $Area B$ spatially, Figure \ref{fig: AndII-ring-pv} presents the position-velocity (PV) diagrams for the open ring-like region of $Area A$. The selected PV diagram paths, highlighted in red in the left panel of Figure \ref{fig: AndII-ring-pv}, indicate the initial direction with arrows. Path 1 corresponds to the area of the stellar stream discovered by \citet{amorisco2014remnant}, while path 2 represents the estimated gas region of $Area A$. There is a gap around the velocity of $-145$ \kms, a fraction of the gas component may fall in the velocity range between $-145$\kms and $-190$ \kms are associated with Andromeda II and the components below $-145$ \kms are from the Milky Way. 

The PV diagrams from $Area A$ to $Area B$ are presented in Figure~\ref{fig: AndII-lines-pv}. Three parallel PV diagrams were constructed to explore the connectivity between the gas structures surrounding Andromeda II and the larger gas formations in its periphery at various locations. At the position of  $32^\prime$ along Path 3, an indistinct boundary between two distinct gas structures. Along Path 4, the gas divides into two segments around the $20^\prime$ mark, with additional low-column density gas clusters detected within $10^\prime$. Along Path 5, the initial boundary emerges at $17^\prime$, followed by a subsequent boundary at $25^\prime$. The structures beyond the position of $25^\prime$ have a lower velocity, and also a lower velocity dispersion (See moment 2 in Fig.~\ref{fig: AndII-M012}), likely originating from high-velocity gas clouds in the Milky Way.  

The PV diagram along the circle paths (Fig.\ref{fig: AndII-ring-pv}) and alone the Area A and B (Fig.\ref{fig: AndII-lines-pv}\added{)} all indicate that there are a few structures lying along a C-shape arc in Area A, but separated from the Milky Way components in Area B. Furthermore, the averaged line-of-sight velocities of the stellar stream near Andromeda II are overlaid on the right panels of Figure~\ref{fig: AndII-ring-pv} and Figure~\ref{fig: AndII-lines-pv} with black dashed lines, revealing a significant agreement between these stellar velocities and those observed within the gaseous $Area A$. Therefore, it is very possible that the c-shape arc structure in Area A is associated with the CGM of the galaxy Andromeda II. 

\subsection{NGC 205} \label{subsec:205-results}

\begin{figure*}[ht!]
\includegraphics[width=\textwidth, angle=0]{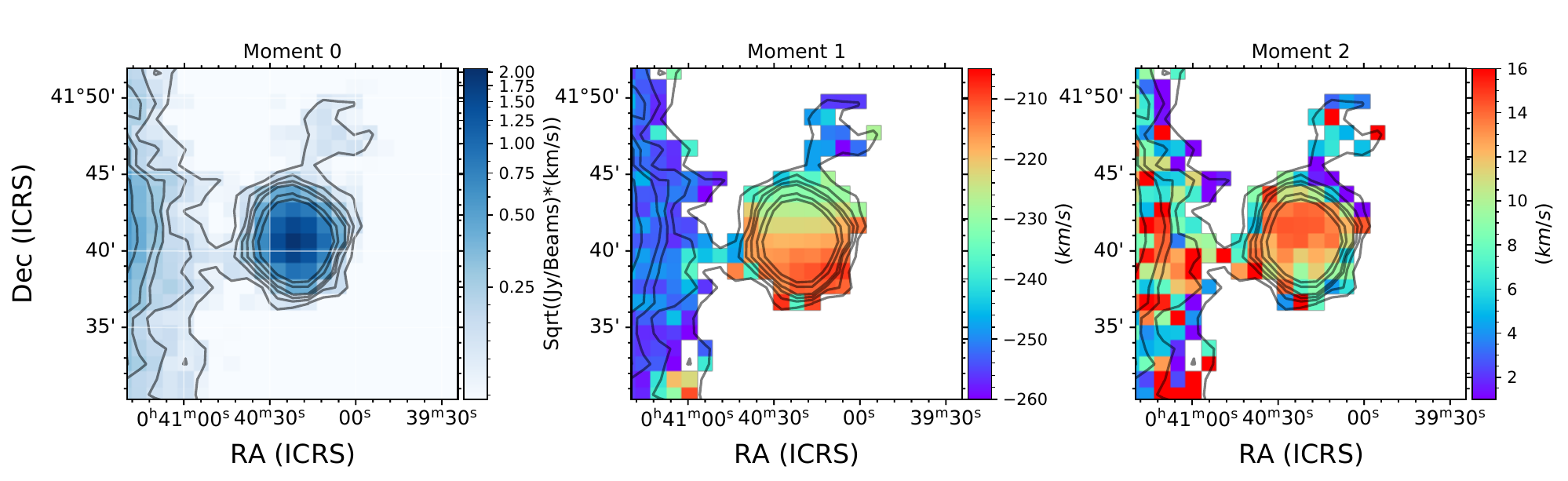}
\caption{The Moment 0, 1, 2 of NGC 205 with velocity range from $-260$ \kms to -185 \kms, the contours illustrate the \HI column density levels of 4, 6, 8, 10, 15, 20, and 50 $\times 10^{18}$ \cms. An extended gaseous structure emerges to the north of NGC 205, extending in a direction away from M31. }
\label{fig:NGC205-m0}
\end{figure*}

\begin{figure}[ht!]
\includegraphics[width=8cm, angle=0]{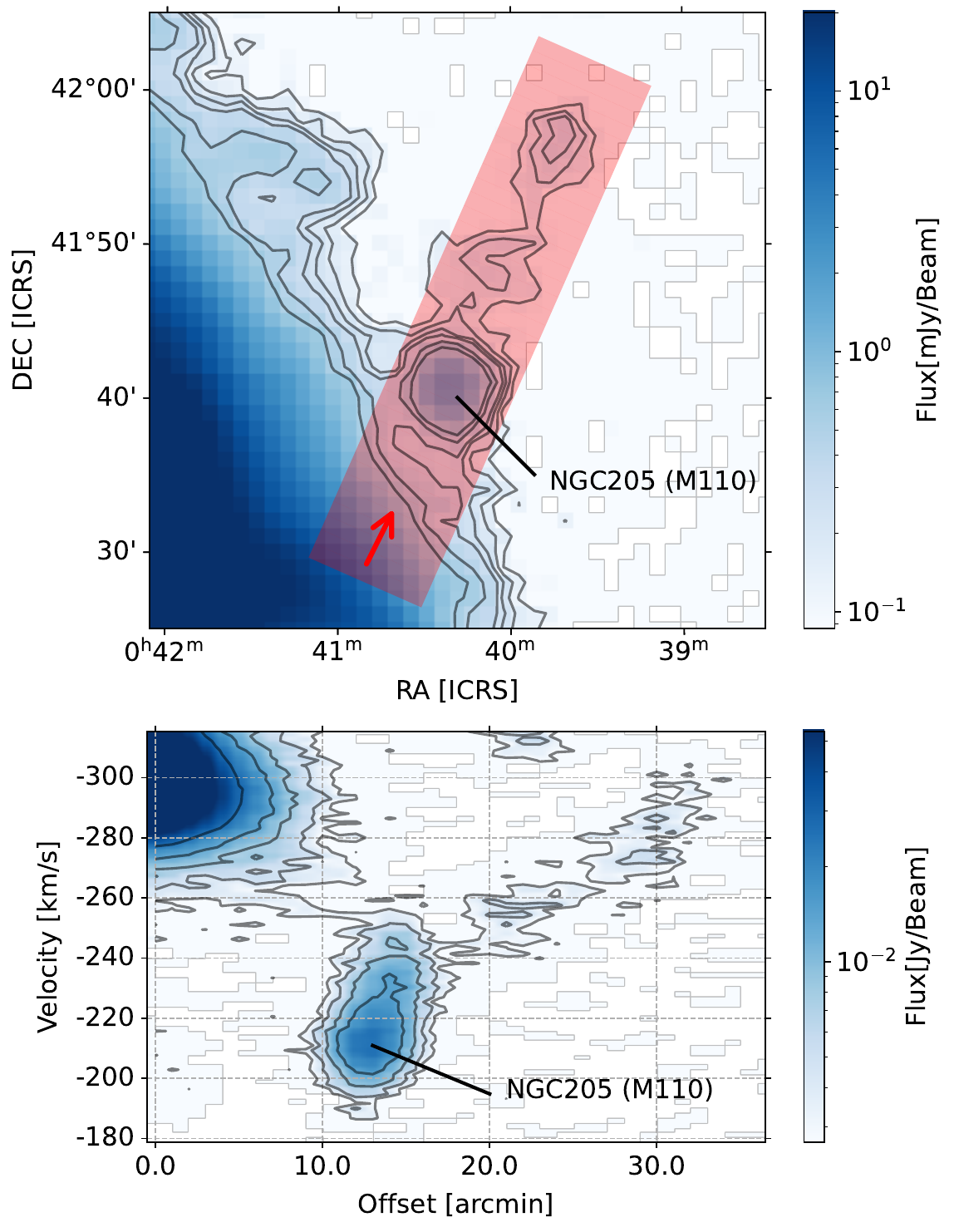}
\caption{Top: The Moment 0 of NGC205 with velocity range from $-305$ \kms to $-180$ \kms, the contours illustrate the \HI column density levels of 4, 6, 8, 10, 15, 20, and 50 $\times 10^{18}$ \cms. Bottom: The position and velocity diagram of the red path line in the top figure, contours highlight the 3, 4, and 5 $\sigma$ limits. A \HI tail structure shows in the northeast direction of NGC205, pointing at a higher velocity region.}
\label{fig: NGC205-pv-mini}
\end{figure}

\begin{figure*}[ht!]
\includegraphics[width=\textwidth, angle=0]{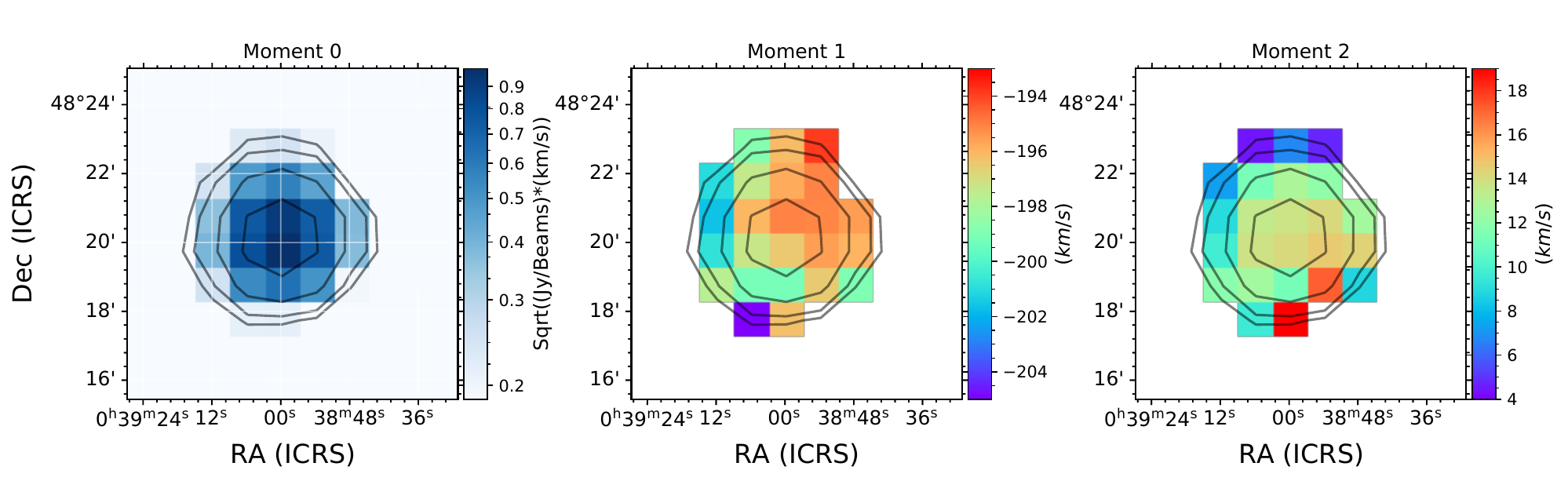}
\caption{The moment 0, 1, 2 of NGC 185 from $-260$ \kms to $-185$ \kms, the contours illustrate the \HI column density levels of 6.5, 9, 15, and 20 $\times 10^{18} $ \cms.}
\label{fig: NGC185-m0-hipro}
\end{figure*}

The Moments 0, 1, and 2 maps of NGC 205 are presented in Figure \ref{fig:NGC205-m0} individually. In the calculation of Moment 0, the velocity range considered is restricted between $-260$ \kms and $-185$ \kms. Contour levels representing the \HI column density are set at values of 4, 6, 8, 10, 15, 20, and 50 $\times 10^{18} $ \cms. The centroid velocity of the \HI profile of NGC 205 is measured at $-221.224$ \kms, with an observed integrated \HI flux of 2.815 Jy \kms. Given the galaxy's distance of 0.80 Mpc, the calculated \HI mass amounts to $(4.2\pm0.3) \times 10^{5} M_{\odot}$. Notably, due to the close proximity of NGC 205 to its host galaxy M31, some of the M31 gas within the same velocity range is included in the Moment 0 map. This inclusion results in an apparent connection between M31 and the western side of NGC 205 at the outermost contour. However, an examination of Moment 1 indicates that the structures of the two galaxies are not directly linked. The PV diagrams of NGC 205 presented in Figure~\ref{fig: NGC205-pv-mini} further demonstrate that the gas structures of M31 and NGC 205 remain distinct and have not undergone merging. Some signals originating from M31 may overlap with the spectral signature of NGC 205, potentially affecting the morphology of the extended wing in the overall \HI profile at higher velocities. To address this issue, a multi-Gaussian fitting method is applied to subtract the \HI components originating from the Milky Way. 

Moreover, in the northern region of NGC 205 in Figure \ref{fig:NGC205-m0}, extended gas structures are observed moving away from M31. The Moment 1 map also shows that the velocities on the northern side of NGC 205 are similar to those in the extended region. To further explore the extended gas structure north of NGC 205, a Moment 0 map is generated over a broader velocity range from $-305$ \kms to $-180$ \kms, covering NGC 205 and its surrounding area within 40 arc minutes, as depicted in Figure \ref{fig: NGC205-pv-mini}. This investigation reveals an additional gas clump along the direction from M31 to NGC 205 and a tail-like gaseous trail spread across a velocity range from $-240$ \kms to $-310$ \kms. The lower-velocity end of this trail almost connects with NGC 205. 

In relation to the unresolved issue of the missing ISM mass in NGC 205, which involves the discrepancy between theoretical expectations and the actual observed mass of the ISM, one theory suggests that tidal interactions with M31 have resulted in a significant loss of ISM mass \citep{2006AJ....131..332G,2012MNRAS.423.2359D,2016ASSL..420..191F}. A piece of evidence that points to the tidal impact of M31 on NGC 205 is the existence of the NGC 205 Loop, a curved stellar stream \citep{2001Natur.412...49I}, which \cite{2004MNRAS.351L..94M} ascribes to a forerunner of NGC 205, created by tidal interactions between NGC 205 and M31. We have determined that the position of the tail-like \HI structure to the north of NGC 205 coincides with the extended direction of the curved stellar stream near NGC 205. However, the tail gas structure extends towards the high-velocity end, in contrast to the extension towards the low-velocity end observed in the stellar stream. Therefore, additional observational evidence is required to ascertain whether there is a direct correlation between the two occurrences.

\subsection{NGC 185} \label{subsec:185-results}

Figure \ref{fig: NGC185-m0-hipro} presents the Moment 0, 1, and 2 maps of NGC 185 individually. The \HI column density contours are shown at levels of 6.5, 9, 15, 20, and 25 $\times 10^{18}$ \cms, progressively from the outer to the inner areas. The overall \HI profile displays a central velocity of $-200.875$ \kms. In the Moment 0, the velocity range is from $-235$ \kms to $-130$ \kms. However, there was no \HI emission detected beyond the velocity limits of $-230$ to $-140$ \kms. The total \HI flux identified within NGC 185 is 0.997 Jy \kms, at a distance of 0.65 Mpc, corresponding to an \HI mass of $(1.0\pm0.1)\times10^{5} M_{\odot}$.
It is anticipated that the ISM within NGC 185 will be primarily concentrated in its current star formation regions, displaying morphology and kinematic characteristics that align with its internal stellar population \citep{1996ApJ...470..781W,1997ApJ...476..127Y,1999AJ....118.2229M,2001AJ....122.1747Y}. Our findings are in agreement with prior studies, as we did not identify any extra gas structures around NGC 185 that could be logically linked to it. NGC 185 has a comparable stellar formation and evolutionary history to NGC 205, and it also suffers from an ISM shortage \citep{1998ApJ...507..726S,2006MNRAS.369.1321D}. However, the lack of evident interaction signatures makes it difficult to attribute the ISM deficit of NGC 185 solely to M31's tidal stripping. The ISM deficiency in NGC 185 may also be influenced by supernova feedback, as indicated by the existence of supernova remnants in this dwarf galaxy \citep{1984ApJ...281L..63G,1997ApJ...476..127Y,1998ARA&A..36..435M,2007AJ....134.2148L}.

\section{Conclusion} \label{sec:conc}
This study showcases the findings from FAST observations regarding the \HI contents within and surrounding three M31 satellite galaxies: Andromeda II, NGC 185, and NGC 205. Notably, we discovered certain HI structures around And II and NGC 205. These findings illuminate the unresolved issue of their missing ISM and contribute to our understanding of their interactions with their host, the M31 galaxy.

Andromeda II exhibits a notable absence of \HI emission at its core, with a concentration primarily observed in a c-shape structure identified as $Area A$ along its periphery. The spatial alignment of $Area A$ bears a remarkable resemblance to the stellar stream encircling Andromeda II, suggesting lingering effects from tidal interactions. Examination of kinematics through PV diagrams and moment maps indicates distinct velocity characteristics between the gaseous envelopes labeled as $Area A$ and outer regions, potentially associated with Andromeda II.

Regarding NGC 185 and NGC 205,  the \HI content of NGC 185 results in an integrated \HI flux leading to an estimated \HI mass of $(1.0\pm0.1) \times 10^{5} M_{\odot}$. In contrast, the total \HI mass of NGC 205 is calculated to be $(4.2\pm0.3) \times10^{5} M_{\odot}$. Variations in velocities observed through Moment 1 and PV diagrams challenge apparent gas connections to M31, effectively distinguishing the gas structures of the two galaxies. Furthermore, the extension of \HI tidal features from NGC 205 towards the northeast direction implies complex galactic interactions within the intergalactic medium. However, these components may be helpful to resolve the missing ISM problem for NGC 205. 

These thorough observations offer a significant understanding of the dynamics within the Local Group, with a particular emphasis on the composition of gases and movement patterns around Andromeda's satellite galaxies. These dynamics shed light on the intricate interplays between galaxies and the space between galaxies, establishing a foundation for modeling the development of dwarf satellites as a result of these interplays. Moreover, examining the disparities between gas and star elements could uncover crucial information about the role of gravity in the evolution of galaxies, contributing to enhancements in simulations of the Local Group's historical dynamics and predicting the fate of dispersed interstellar matter within the wider cosmic framework.

\begin{acknowledgements}
We would like to express our gratitude to the referee for the valuable comments. We acknowledge the support of the China National Key Program for Science and Technology Research and Development of China (2022YFA1602901), the National Natural Science Foundation of China (Nos. 11988101, 11873051), the CAS Project for Young Scientists in Basic Research Grant (No. YSBR-062), and the K.C. Wong Education Foundation, and the science research grants from the China Manned Space Project. Y Jing acknowledges support from the Cultivation Project for FAST Scientific Payoff and Research Achievement of CAMS-CAS.

This work has used the data from the Five-hundred-meter Aperture Spherical Radio Telescope ( FAST ). FAST is a Chinese national mega-science facility, operated by the National Astronomical Observatories of the Chinese Academy of Sciences (NAOC).
\end{acknowledgements}

\bibliographystyle{raa}
\bibliography{ms2024-0118}

\begin{thebibliography}{38}
\providecommand\natexlab[1]{#1}
\providecommand\JournalTitle[1]{#1}

\bibitem[Amorisco {et~al.}(2014)]{amorisco2014remnant}
Amorisco, N.~C., Evans, N.~W., \& van~de Ven, G. 2014, Nature, 507, 335

\bibitem[{Bender} {et~al.}(1991)]{1991A&A...246..349B}
{Bender}, R., {Paquet}, A., \& {Nieto}, J.~L. 1991, \aap, 246, 349

\bibitem[{Blitz} \& {Robishaw}(2000)]{2000ApJ...541..675B}
{Blitz}, L., \& {Robishaw}, T. 2000, \apj, 541, 675

\bibitem[{Cepa} \& {Beckman}(1988)]{1988A&A...200...21C}
{Cepa}, J., \& {Beckman}, J.~E. 1988, \aap, 200, 21

\bibitem[{Davidge}(2005)]{2005AJ....130.2087D}
{Davidge}, T.~J. 2005, \aj, 130, 2087

\bibitem[{De Looze} {et~al.}(2012)]{2012MNRAS.423.2359D}
{De Looze}, I., {Baes}, M., {Parkin}, T.~J., {et~al.} 2012, \mnras, 423, 2359

\bibitem[{De Looze} {et~al.}(2017)]{2017MNRAS.465.3741D}
{De Looze}, I., {Baes}, M., {Cormier}, D., {et~al.} 2017, \mnras, 465, 3741

\bibitem[{De Rijcke} {et~al.}(2006)]{2006MNRAS.369.1321D}
{De Rijcke}, S., {Prugniel}, P., {Simien}, F., \& {Dejonghe}, H. 2006, \mnras, 369, 1321

\bibitem[{del Pino} {et~al.}(2017)]{2017MNRAS.469.4999D}
{del Pino}, A., {{\L}okas}, E.~L., {Hidalgo}, S.~L., \& {Fouquet}, S. 2017, \mnras, 469, 4999

\bibitem[{Ebrov{\'a}} {et~al.}(2019)]{2019IAUS..344...62E}
{Ebrov{\'a}}, I., {{\L}okas}, E.~L., {Fouquet}, S., \& {Del Pino}, A. 2019, in Dwarf Galaxies: From the Deep Universe to the Present, ed. K.~B.~W. {McQuinn} \& S.~{Stierwalt}, Vol. 344, 62

\bibitem[{Einasto} {et~al.}(1974)]{1974Natur.252..111E}
{Einasto}, J., {Saar}, E., {Kaasik}, A., \& {Chernin}, A.~D. 1974, \nat, 252, 111

\bibitem[{Ferguson} \& {Mackey}(2016)]{2016ASSL..420..191F}
{Ferguson}, A. M.~N., \& {Mackey}, A.~D. 2016, in Astrophysics and Space Science Library, Vol. 420, Tidal Streams in the Local Group and Beyond, ed. H.~J. {Newberg} \& J.~L. {Carlin}, 191

\bibitem[{Fouquet} {et~al.}(2017)]{2017MNRAS.464.2717F}
{Fouquet}, S., {{\L}okas}, E.~L., {del Pino}, A., \& {Ebrov{\'a}}, I. 2017, \mnras, 464, 2717

\bibitem[{Gallagher} {et~al.}(1984)]{1984ApJ...281L..63G}
{Gallagher}, J.~S., I., {Hunter}, D.~A., \& {Mould}, J. 1984, \apjl, 281, L63

\bibitem[{Geha} {et~al.}(2006)]{2006AJ....131..332G}
{Geha}, M., {Guhathakurta}, P., {Rich}, R.~M., \& {Cooper}, M.~C. 2006, \aj, 131, 332

\bibitem[{Grcevich} \& {Putman}(2009)]{2009ApJ...696..385G}
{Grcevich}, J., \& {Putman}, M.~E. 2009, \apj, 696, 385

\bibitem[{Grcevich} \& {Putman}(2016)]{2016ApJ...824..151G}
{Grcevich}, J., \& {Putman}, M.~E. 2016, \apj, 824, 151

\bibitem[{Grebel} {et~al.}(2003)]{2003AJ....125.1926G}
{Grebel}, E.~K., {Gallagher}, John~S., I., \& {Harbeck}, D. 2003, \aj, 125, 1926

\bibitem[{Ho} {et~al.}(2012)]{2012ApJ...758..124H}
{Ho}, N., {Geha}, M., {Munoz}, R.~R., {et~al.} 2012, \apj, 758, 124

\bibitem[{Ibata} {et~al.}(2001)]{2001Natur.412...49I}
{Ibata}, R., {Irwin}, M., {Lewis}, G., {Ferguson}, A. M.~N., \& {Tanvir}, N. 2001, \nat, 412, 49

\bibitem[{Jiang} {et~al.}(2019)]{2019SCPMA..6259502J}
{Jiang}, P., {Yue}, Y., {Gan}, H., {et~al.} 2019, Science China Physics, Mechanics, and Astronomy, 62, 959502

\bibitem[{Jiang} {et~al.}(2020)]{2020RAA....20...64J}
{Jiang}, P., {Tang}, N.-Y., {Hou}, L.-G., {et~al.} 2020, Research in Astronomy and Astrophysics, 20, 064

\bibitem[{Jing} {et~al.}(2024)]{2024arXiv240117364J}
{Jing}, Y., {Wang}, J., {Xu}, C., {et~al.} 2024, Science China Physics, Mechanics, and Astronomy, 67, 259514

\bibitem[{Karachentsev} {et~al.}(2018)]{2018MNRAS.479.4136K}
{Karachentsev}, I.~D., {Kaisina}, E.~I., \& {Makarov}, D.~I. 2018, \mnras, 479, 4136

\bibitem[{Lokas} {et~al.}(2014)]{2014MNRAS.445L...6L}
{Lokas}, E.~L., {Ebrova}, I., {Del Pino}, A., \& {Semczuk}, M. 2014, \mnras, 445, L6

\bibitem[{Lucero} \& {Young}(2007)]{2007AJ....134.2148L}
{Lucero}, D.~M., \& {Young}, L.~M. 2007, \aj, 134, 2148

\bibitem[{Mart{\'\i}nez-Delgado} {et~al.}(1999)]{1999AJ....118.2229M}
{Mart{\'\i}nez-Delgado}, D., {Aparicio}, A., \& {Gallart}, C. 1999, \aj, 118, 2229

\bibitem[{Mateo}(1998)]{1998ARA&A..36..435M}
{Mateo}, M.~L. 1998, \araa, 36, 435

\bibitem[{McConnachie}(2012)]{2012AJ....144....4M}
{McConnachie}, A.~W. 2012, \aj, 144, 4

\bibitem[{McConnachie} {et~al.}(2004)]{2004MNRAS.351L..94M}
{McConnachie}, A.~W., {Irwin}, M.~J., {Lewis}, G.~F., {et~al.} 2004, \mnras, 351, L94

\bibitem[{Putman} {et~al.}(2021)]{2021ApJ...913...53P}
{Putman}, M.~E., {Zheng}, Y., {Price-Whelan}, A.~M., {et~al.} 2021, \apj, 913, 53

\bibitem[{Sage} {et~al.}(1998)]{1998ApJ...507..726S}
{Sage}, L.~J., {Welch}, G.~A., \& {Mitchell}, G.~F. 1998, \apj, 507, 726

\bibitem[{Simon}(2019)]{2019ARA&A..57..375S}
{Simon}, J.~D. 2019, \araa, 57, 375

\bibitem[{Spekkens} {et~al.}(2014)]{2014ApJ...795L...5S}
{Spekkens}, K., {Urbancic}, N., {Mason}, B.~S., {Willman}, B., \& {Aguirre}, J.~E. 2014, \apjl, 795, L5

\bibitem[{Thilker} {et~al.}(2005)]{2005ASPC..331..113T}
{Thilker}, D.~A., {Braun}, R., \& {Westmeier}, T. 2005, in Astronomical Society of the Pacific Conference Series, Vol. 331, Extra-Planar Gas, ed. R.~{Braun}, 113

\bibitem[{Welch} {et~al.}(1996)]{1996ApJ...470..781W}
{Welch}, G.~A., {Mitchell}, G.~F., \& {Yi}, S. 1996, \apj, 470, 781

\bibitem[{Young}(2001)]{2001AJ....122.1747Y}
{Young}, L.~M. 2001, \aj, 122, 1747

\bibitem[{Young} \& {Lo}(1997)]{1997ApJ...476..127Y}
{Young}, L.~M., \& {Lo}, K.~Y. 1997, \apj, 476, 127

\end{thebibliography}
\label{lastpage}
\end{document}